\documentclass[sn-mathphys]{sn-jnl}

\jyear{2021}%

\theoremstyle{thmstyleone}%
\theoremstyle{thmstyletwo}%

\theoremstyle{thmstylethree}%

\usepackage{graphicx}
\usepackage{booktabs}
\usepackage{multirow}
\usepackage{amssymb}
\usepackage{amsmath}
\usepackage{color}
\usepackage{pifont}
\usepackage{tabularx}
\usepackage{graphics}
\usepackage{makecell}
\usepackage{balance}
\usepackage{subcaption}
\usepackage{hyperref}
\usepackage{textcomp}
\usepackage{comment}
\usepackage{xcolor}
\definecolor{light-gray}{gray}{0.95}

\newcommand{\lab}[1]{\textquotesingle{#1}\textquotesingle}
\usepackage{xcomment}
\usepackage{adjustbox}

\usepackage{xspace}
\newcommand{\tool}[1]{\textsc{#1}\xspace}

\newcommand{\dnncov}{\tool{DNNCov}}
\usepackage{listings, xcolor}
\usepackage{array}

\usepackage{booktabs}

\usepackage{verbatim}
\usepackage{array}
\newcommand{\PreserveBackslash}[1]{\let\temp=\\#1\let\\=\temp}

\newcolumntype{C}[1]{>{\PreserveBackslash\centering}p{#1}}
\newcolumntype{R}[1]{>{\PreserveBackslash\raggedleft}p{#1}}
\newcolumntype{L}[1]{>{\PreserveBackslash\raggedright}p{#1}}
\makeatletter
\newcommand{\linebreakand}{%
  \end{@IEEEauthorhalign}
  \hfill\mbox{}\par
  \mbox{}\hfill\begin{@IEEEauthorhalign}
}

\begin{document}

\title[An Overview ...]{An Overview of Structural Coverage Metrics for Testing Neural Networks}

\author[1]{\fnm{Muhammad} \sur{Usman}}

\author[2]{\fnm{Youcheng} \sur{Sun}}

\author[3]{\fnm{Divya} \sur{Gopinath}}

\author[4]{\fnm{Rishi} \sur{Dange}}

\author[5]{\fnm{Luca} \sur{Manolache}}

\author*[6]{\fnm{Corina S.} \sur{P\u{a}s\u{a}reanu}}\email{pcorina@cmu.edu}
\affil[1]{\orgdiv{University of Texas, Austin}}

\affil[2]{\orgdiv{University of Manchester, UK}}

\affil[3]{\orgdiv{NASA Ames, KBR}}
\affil[4]{\orgdiv{Princeton University}}
\affil[5]{\orgdiv{Palo Alto High School}}
\affil[6]{\orgdiv{NASA Ames, KBR, CMU Cylab}}

\abstract{Deep neural network (DNN) models, including those used in safety-critical domains, need to be thoroughly tested to ensure that they can reliably perform well in different scenarios. In this article, we provide an overview of structural coverage metrics for testing DNN models, including neuron coverage (NC), k-multisection neuron coverage (kMNC), top-k neuron coverage (TKNC), neuron boundary coverage (NBC), strong neuron activation coverage (SNAC) and modified condition/decision coverage (MC/DC). We evaluate the metrics on realistic DNN models used for perception tasks (including  LeNet-1, LeNet-4, LeNet-5, and ResNet20) as well as on networks used in autonomy (TaxiNet).
We also provide a tool, \dnncov, which can measure the testing coverage for all these metrics. \dnncov outputs an informative coverage report to enable researchers and practitioners to assess the adequacy of DNN testing, compare different coverage measures, and to more conveniently inspect the model’s internals during testing.


}

\keywords{Coverage, Neural Networks, Testing}



\maketitle

\section{Introduction}
\label{sec:introduction}
Today’s world has seen a significant rise in deep learning models that are used for solving complex tasks, such as medical diagnosis, image and video recognition, text processing, program understanding, and so on. Since such models are increasingly used for safety-critical applications, including the perception and control of self-driving vehicles, ensuring that deep learning models perform as expected is of extreme importance. 
Typically, statistical accuracy on a set of tests is used as a measure of model performance. However, it is unclear if the tests cover all the possible behaviors of the model, including corner cases. Coverage metrics can be used to measure the adequacy of testing. 
In this article we review and evaluate recently proposed structural coverage metrics for testing neural networks. In addition, we describe a tool, \dnncov, which incorporates into a common framework these different coverage metrics, to allow for easy experimentation and comparison. 


Specifically, we evaluate neuron coverage and its variants, as well as a more complex MC/DC (modified condition/decision coverage) criterion for neural networks. Neuron coverage was first proposed in DeepXplore \cite{pei2017deepxplore} where it was developed  as a DNN counterpart for statement coverage. This was later extended to a family of more fine-grained criteria \cite{ma2018deepgauge}, such as k-multisection neuron coverage (KMNC), top-k neuron coverage (TKNC), top-k neuron patterns (TKNP), neuron boundary coverage (NBC) and strong neuron activation coverage (SNAC) \cite{ma2018deepgauge}. 
In \cite{sun2019structural}, four MC/DC  variants were defined for DNN testing. Subsequently, there has been a boom in coverage-guided DNN testing techniques. ADAPT \cite{10.1145/3395363.3397346} uses an adaptive learning algorithm to generate new images for testing neural networks with an aim of increasing e.g., neuron coverage. DeepHunter~\cite{deephunter} is a fuzzing framework which uses metamorphic mutations to generate tests following the metrics in \cite{ma2018deepgauge} while the concolic testing in \cite{sun2018concolic} targets MC/DC.

Despite the proliferation of the aforementioned approaches, it is still not well understood which technique or criteria can adequately determine if a neural network model has been well tested with respect to properties such as functional diversity or vulnerability to attacks (e.g., adversarial input perturbations) so on. The problem is challenging due to the opaque nature of the networks. The coverage criterion that is the most effective may vary from one model to the other, and from one task to another. 
In this work, we present \dnncov, a unified framework for evaluating, comparing and visualizing different structural coverage metrics for a given model and test suite.

We summarize our contributions as follows:
\begin{itemize}
\item  We provide an overview of different structural coverage metrics (NC, KMNC, TKNC, NBC, SNAC, and MC/DC) for testing neural networks. 
\item We evaluate the criteria with respect to functional diversity and defect detection. We consider both adversarial robustness and data poisoning scenarios (the latter has not been considered in previous evaluations). We consider state-of-the art models trained for classification and regression tasks.
\item We describe a tool, \dnncov, that incorporates multiple structural testing criteria, including the complex MC/DC coverage proposed in \cite{sun2019structural} (not considered in previous evaluations) to
enable easy comparison between different  testing metrics. \dnncov computes multiple coverage criteria simultaneously giving a 2.5x time improvement over the baseline sequential implementation.  
\item \dnncov has a visualization component that displays the neural network architecture along with the coverage achieved (neurons covered or not), and quantitative information (measuring the number of tests that achieve the coverage). We believe that this nuanced view of the coverage, can enable developers to better understand and debug the behavior of the neural networks.
\end{itemize}

\subsection{Applications} \dnncov can be used to compare different coverage metrics on the same model and reason about the effectiveness of a metric in determining test set adequacy. Moreover, one can compare the coverage metrics over different models (for the same task) to assess how different architectures impact testing adequacy. One can also compare different test sets with varying sizes to determine how well they test a given model. Our evaluation (Section~\ref{sec:evaluation}) has experiments highlighting these applications. If a smaller test suite achieves same coverage as that achieved by the full test suite, the test set can also be reduced. Furthermore, \dnncov can enable coverage-driven test generation for different criteria and can be used for optimizing neural networks, for instance by pruning the neurons that are never covered (or have low coverage). 


\subsection{Comparison with Related Studies} There are many approaches and metrics that have been proposed recently, aiming to measure the testing adequacy of DNN models \cite{pei2017deepxplore,ma2018deepgauge,sun2019structural}. There are also frameworks proposed to facilitate such measurements, such as \cite{EvalDNN}. 
In contrast to previous studies, we provide the following contributions. We evaluate different structural metrics with respect to both functional diversity and defect detection. The latter is evaluated not only with respect to adversarial robustness (which is overwhelmingly the only one studied in the related literature), but also with respect to poisoning attacks, which provide additional interesting insights on different coverage metrics. We consider models trained for both classifications and regression tasks. Furthermore, we provide a tool that incorporates these criteria, including the complex MC/DC criterion not considered in previous studies. \dnncov provides finer grained results, as it computes not only qualitative results (as was done also in previous work) but also quantitative information about {\em how many} inputs satisfy the criteria.
\section{Structural Coverage Criteria for Neural Networks}
\label{sec:background}


\subsection{Neural Networks.} Neural networks (NNs) 
are machine learning algorithms that can be trained to perform different tasks such as classification and regression.
NNs consist of multiple layers, starting from the \textit{input} layer, followed by one or more \textit{hidden} layers (such as convolutional, dense, activation, and pooling), and a final \textit{decision} layer.
Each layer consists of a number of computational units, called \textit{neurons}. Each neuron applies an activation function on a weighted sum of its inputs (coming from the previous layer).
$N(X)=\sigma(\sum_i w_i \cdot N_i(X) + b)$
where $N_i$ denotes the value of the $i^{th}$ neuron in the previous layer of the network and the coefficients $w_i$ and the constant $b$ are referred to as \emph{weights} and \emph{bias}, respectively; $\sigma$ represents the activation function.
For instance, the ReLU (rectified linear unit) activation function returns its input as is if it is positive, and returns 0 otherwise, i.e., $\sigma(X) = max(0, X)$.
The final decision layer (\textit{logits}) typically uses a specialized function (e.g., max or \textit{softmax}) to determine the decision or the output of the network.

\subsection{DNN Structural Coverage Metrics}
In this study we evaluate neuron coverage \cite{pei2017deepxplore} and its extensions \cite{ma2018deepgauge} and the MC/DC variants for DNNs \cite{sun2019structural}.
We describe each of these criteria  by formulating their test conditions.
Given a test suite and a DNN model, the coverage for each criterion is thus computed as the portion of test conditions that are satisfied.

In the following, we use $a_{l,i}$ to denote a neuron's activation value, where $l$ is the layer index and $i$ is the neuron index at that layer. 

\noindent\textbf{Neuron Coverage (NC).} NC can be seen as a  statement coverage variant for DNNs. A neuron $n_{l,i}$ is said to be covered, if its neuron activation value ($a_{l,i}$) is larger than 0 (or some specified threshold) at least by one of the test inputs. Thus, the set of test conditions to be met for NC can be formulated as follows, where $L$ is the total number of layers in the DNN.
$$\{a_{l,i}>0, 1<l<L\}$$ 

\noindent\textbf{Neuron Boundary Coverage (NBC).} NBC extends NC by considering the neuron activations at the maximum and minimum boundary cases. Assuming $high_{l,i}$ and $low_{l,i}$ are respectively the estimated lower and upper bounds on the neuron activation $a_{l,i}$'s value, then we can formulate the set of test conditions for NBC as follows.
$$\{a_{l,i}>high_{l,i}, a_{l,i}<low_{l,i}, 1<l<L\}$$ 
The estimation of lower/upper bounds is typically done via profiling the training dataset. 

\noindent\textbf{Strong Neuron Activation Coverage (SNAC).} SNAC focuses on test conditions on  corner cases with respect to the upper boundary value.  $$\{a_{l,i}>high_{l,i}, 1< l< L\}$$

\noindent\textbf{K-Multisection Neuron Coverage (KMNC).}
KMNC divides a neuron's activation range between $high_{l,i}$ and $low_{l,i}$ into $K$ equivalent sections, each denoted by $range_{l,i,k}$, and test conditions in KMNC are defined as the coverage of these activation sections.
$$\{a_{l,i}\in range_{l,i,k}, 1< l< L, 1\leq k\leq K\}$$

\noindent\textbf{Top-K Neuron Coverage (TKNC).}
Given a test input $x$, a neuron is TKNC covered if its neuron activation value occurs to be one of the most active $K$ neurons at its layer, denoted by $a_{l,i}\in top_K(l,x)$. Thus, the test conditions are as follows.

$$\{a_{l,i}\in top_K(l,x), 1< l< L\}$$

\noindent\textbf{Modified Condition/Decision Coverage (MC/DC).} MC/DC was originally developed by NASA and has been widely used for testing high integrity software. It was recently adapted for testing DNNs. Different from the coverage criteria above, MC/DC takes into account the relation between neuron activations at two adjacent layers, such that its test conditions require that any neuron activation at layer $l+1$ (decision) must be independently impacted by each neuron at layer $l$ (condition).

$$\{\forall i,j,h, change(a_{l,i})\wedge change (a_{l+1,j})\wedge\neg change(a_{l,h}), 1< l< L-1\}$$ 

A Sign (S) change function and a Value (V) change function are defined in \cite{sun2019structural} for depicting how a neuron activation changes when the input changes from a test to another. As a result, there is a family of four variants of MC/DC for DNNs, including SS coverage (SS), SV coverage (SV), VS coverage (VS) and VV coverage (VV). 
To ease the use of MC/DC in large DNN models, in the later evaluation, we generalize its test conditions from single neurons to sets of neurons (feature maps) between two adjacent layers.

In this study we focus on structural coverage criteria for DNNs, due to their popularity, wide use, and similarity to established metrics for general-purpose software. These criteria are thus more likely to be integrated in certification procedures for safety critical systems that contain neural networks. 
There are also other, non-structural proposals for measuring the adequacy of testing DNNs e.g., the safety coverage \cite{huang2017safety} and the surprise adequacy \cite{kim2019guiding}. 
\section{The \dnncov Tool}
\label{sec:tool}
\begin{figure*}
\centering
\includegraphics[scale=0.35]{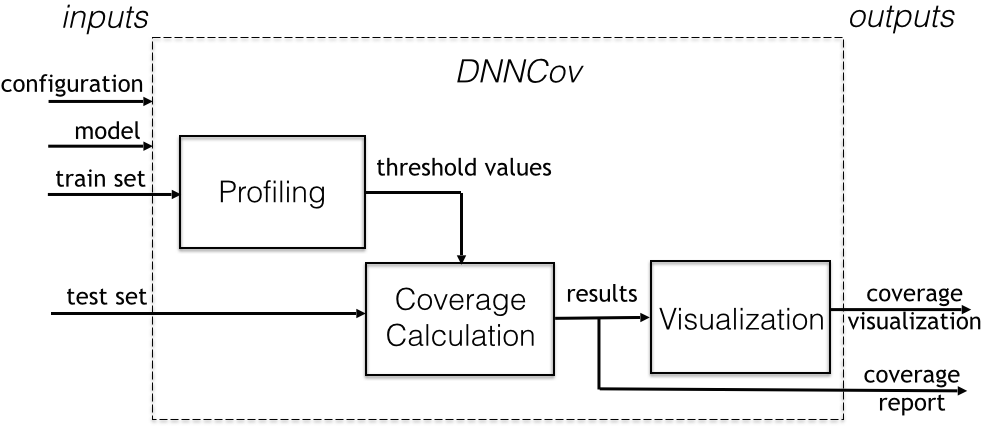}
\caption{The \dnncov Tool.}
\label{fig:framework}
\end{figure*}

\dnncov  (Figure~\ref{fig:framework}) provides an integrated framework for testing neural network models. 
%
The framework takes in an input test set and optionally a training set and computes the different coverage results achieved by the set. The user also needs to 
specify a configuration (\# of training/test  inputs to be used and criteria to calculate). \dnncov profiles the neural network model to calculate threshold values based on the training set. These threshold values are used as input to the coverage calculation module. The coverage calculation module computes coverage for the different metrics. \dnncov then produces a summarized coverage report and a visualization of the results.

\subsection{Implementation, Profiling and Coverage Calculation}
\begin{figure*}
\centering
\includegraphics[scale=0.5]{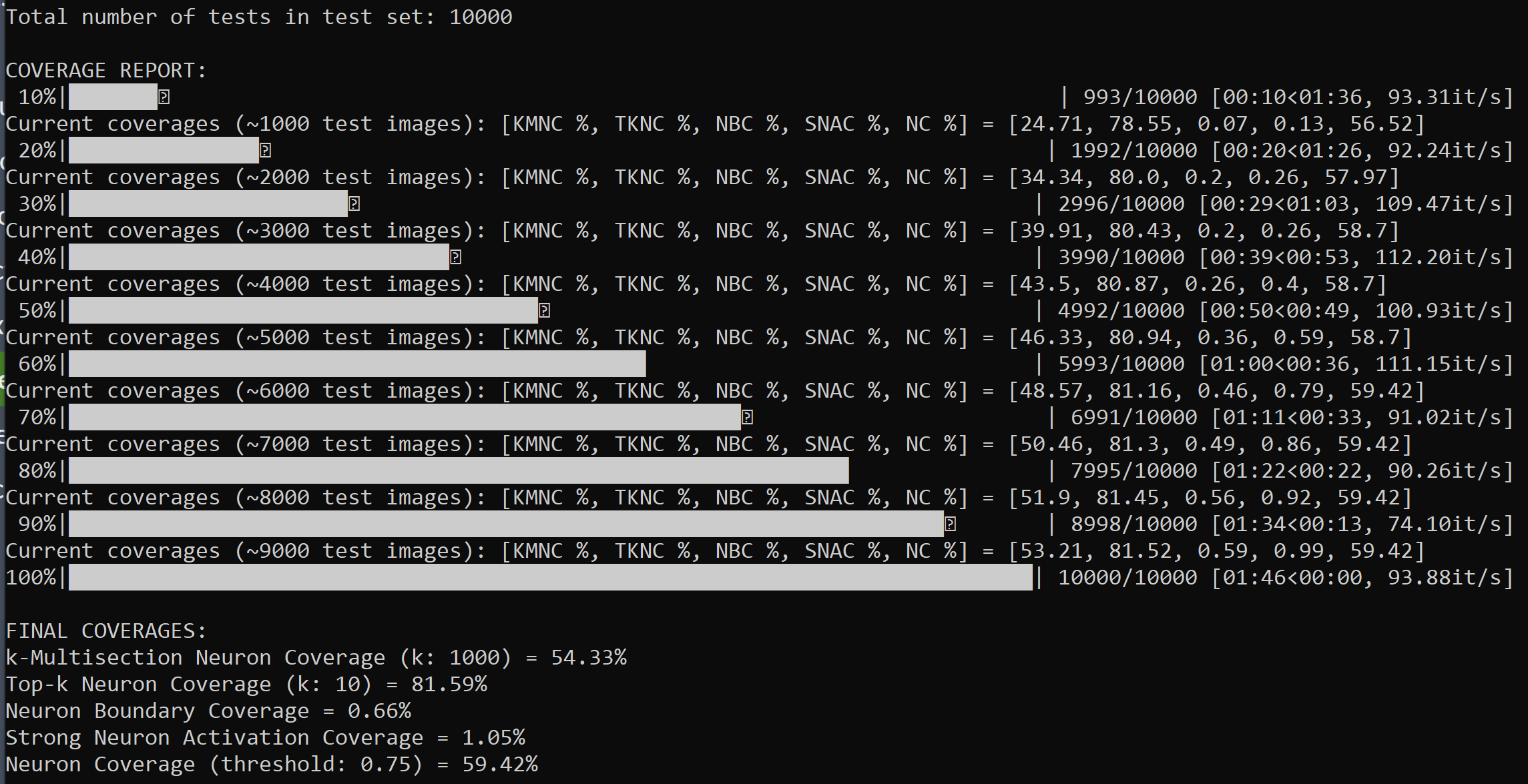}
\caption{Tool Progress on MNIST (LeNet-1)}
\label{fig:progress}
\end{figure*}

To implement \dnncov, we leveraged the DeepHunter \cite{deephunter} codebase, which already provided support for the neuron-based coverage metrics. We extended it with the four MC/DC variants i.e., Sign-Sign (SS), Sign-Value (SV), Value-Sign (VS) and Value-Value (VV). 
We  profile the neural network model to estimate the threshold values needed in some coverage criteria. We wrote new functions to measure coverage for a given test suite, replacing the original test-generation process. We also implemented simultaneous coverage computations to improve efficiency (achieving 2.5x runtime improvement).  
\dnncov outputs an informative coverage report to the user as illustrated in Figure~\ref{fig:progress}.

\subsection{Reporting Quantitative Information}
\dnncov computes not only which coverage obligations are fulfilled, but also records the number of tests that achieve the coverage. 
This quantitative information gives us an idea of which parts of the network have been exercised lesser than others, thus enabling us to direct testing / debugging efforts in those parts. This is specifically important for neural networks since unlike traditional programs, every test exercises almost every neuron of the network and the changes in behavior is due to the different output values of the neurons. Therefore executing a neuron only once or just few times may not suffice to exercise different behaviors or expose vulnerabilities. Different test suites may have similar total coverage values but different coverage distributions. This cannot be highlighted by existing coverage metrics. We calculate minimum, maximum, average, standard deviation and variance of the number of inputs that cover each of the coverage obligations for each coverage criterion.

\subsection{Visualization}

\begin{figure*}
\centering
\includegraphics[scale=0.6]{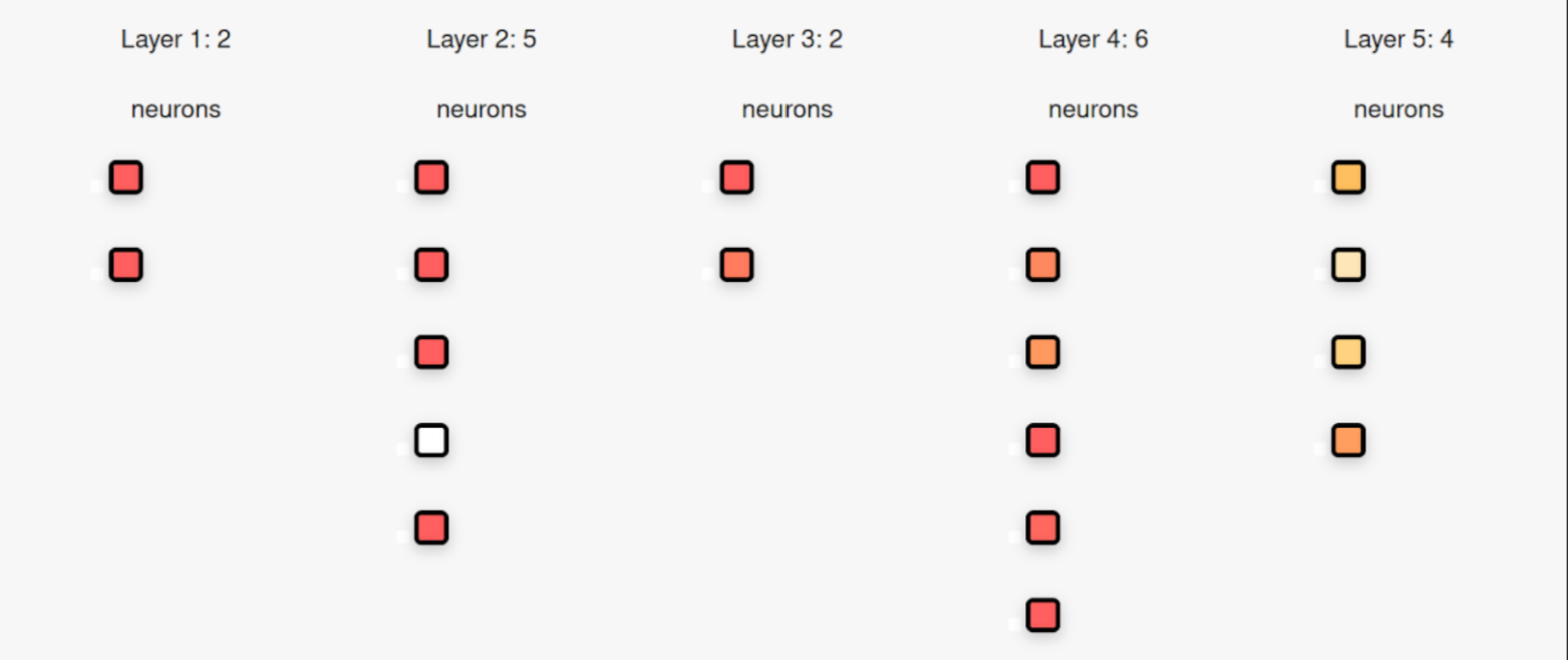}
\caption{Visualization of Neuron Coverage}
\label{fig:visualization}
\end{figure*}

The coverage results computed by \dnncov are saved in a file and are viewed by the visualization component of the tool (which is built based on Tensorflow Playground). This component has a web-based interface that shows the architecture of the model and quantitative coverage information. The user can navigate through the model, and visualize specific information related to the coverage. Neurons are given a color from white (least covered) to red (most covered) based on how many times they are covered (see Figure~\ref{fig:visualization}). 
Neuron pairs covered according to MC/DC criteria are shown as connected with each other using connection lines. The displayed information can be used to identify poorly tested areas in the neural network.

\subsection{Tool Configuration}

Users can load their own trained neural network model using the \lstinline{--model} option. Users can select the coverage criteria by using the \lstinline{--criteria} option. There are three possible settings; \lstinline{all} calculates all six coverage criteria, \lstinline{no-mcdc} excludes MC/DC and \lstinline{mcdc} only calculates MC/DC.

The reason MC/DC is separate is that it works on input pairs whereas the rest of the metrics are computed for each input. Therefore, calculations of MC/DC and the other NC variants are implemented with two seperate modules in \dnncov.
All output results (including the intermediate files used for visualization) are stored in an output folder. Users can specify the directory of the output folder using the \lstinline{--outputs} option. 
Three common datasets can be automatically loaded in \dnncov: \\
\lstinline{--mnist-dataset} for MNIST,\\ \lstinline{--cifar10-dataset} for CIFAR-10 and\\ \lstinline{--tinytaxinet-dataset} for TaxiNet.\\ 
Users' own datasets can be used by specifying the \lstinline{--input-tests} and \lstinline{--input-train} folders. The tool also supports calculation over a subset of training and test dataset. Users can specify the number of training/test inputs to be used by using the \lstinline{--train} and \lstinline{--test} options. 
For example,\\ 
\lstinline{python dnncov.py --criteria all --model lenet1.h5 --outputs outs1}\\ \lstinline{--mnist-dataset --train 60000 --test 10000}\\ calculates coverage for all six criteria for the ~\lstinline{lenet1.h5} model using the default training and test sets for MNIST. 

\section{Evaluation}
\label{sec:evaluation}

We structure our evaluation along the following thrusts.
\begin{itemize}
    \item Comparison of coverage metrics on standard test sets.
    \item Evaluation of coverage information with respect to sensitivity to functional diversity.
    \item Evaluation of coverage information with respect to sensitivity to defect detection; we consider both adversarial and poisoned inputs.
    \item Evaluation of quantitative and visualization information.
\end{itemize}

In turn, this information can be used by developers to select among multiple test sets, favoring the ones for which multiple metrics yield high coverage. Furthermore, developers can use the coverage information to even choose between models (trained for the same task), favoring the model that again achieves high coverage for the majority of metrics.


\noindent{\bf Benchmarks.} 
Table~\ref{tab:subjects} shows the details of the datasets and models used in our study. We used a set of \textit{image classification models} on benchmark datasets. \textbf{MNIST} (Modified National Institute of Standards and Technology database)~\cite{MNIST} is a collection of handwritten digits from \lab{0} to \lab{9}. It has a training set (\emph{MNIST-Train}) with 60k inputs and a test set (\emph{MNIST-Test}) of 10k inputs, which are 28$\times$28 grey scale images. We used the popular \textit{LeNet} convolutional neural network model trained for image classification on the MNIST dataset~\cite{LENET1,LENET2}. \textbf{CIFAR-10} (Canadian Institute For Advanced Research)~\cite{CIFAR-10} is a collection of color images classified to one of 10 classes which include vehicles such as \lab{airplane}, \lab{truck} so on  and animals such as bird, cat so on. The training set (\emph{CIFAR-Train}) contains 50k 32$\times$32 color images and a test set (\emph{CIFAR-Test}) of 10k images. We used the state-of-the-art ResNet20 model for image classification on this dataset~\cite{RESNET20}. 
\begin{table*}
\caption{Details of the Datasets and Neural Network Models.}
\label{tab:subjects}
\centering
\scalebox{0.65}{
\begin{tabular}{lcccc}\toprule
\multirow{2}{*}{\textbf{Model}} & \multirow{2}{*}{\textbf{Benchmark}} & \textbf{\#Dataset} & \textbf{Test} & \textbf{Model} \\ && \textbf{(training,test)}&\textbf{Accuracy}&\textbf{Architecture}\\\hline
LeNet-1 & MNIST & (60k,10k) & 90.6\%&2 conv/2 maxpool \\ \hline
LeNet-4 & MNIST & (60k,10k) & 89.9\% &2 conv/2 maxpool/1 dense\\ \hline
LeNet-5 & MNIST & (60k,10k) & 92.8\% &2 conv/2 maxpool/2 dense\\ \hline
MNIST-Funct. & MNIST & (60k,10k) & 96.34\% &2 conv/4 act/1 maxpool/2 dense/1 flatten\\ \hline
MNIST-Pois. & MNIST & (60k,10k) & Test:98.63, Pois.:10.38\% &2 conv/4 act/1 maxpool/2 dense/1 flatten\\ \hline
MNIST-Adv.& MNIST & (60k,10k) & Test: 97.87\%, Adv.: 28.37\% &2 conv/4 act/1 maxpool/2 dense/1 flatten\\ \hline

\multirow{2}{*}{ResNet20} & \multirow{2}{*}{CIFAR-10} & \multirow{2}{*}{(50k,10k)} & \multirow{2}{*}{68.7\%} &21 conv/19 batchnorm/19 act/9 add\\ 
&&&& /1 globalavgpool\\ \hline
TinyTaxiNet & TaxiNet & (51462,7386) & MAE (CTE:1.44, HE:2.75) &3 dense\\
\hline
\end{tabular}
}
\end{table*}

We also applied \dnncov on a regression model from the autonomy domain, namely \textbf{TaxiNet}, which is a perception model for center-line tracking in airport runways \cite{TinyTaxiNet}. The model takes images of the runway as input and produces two outputs, cross-track (CTE) and heading angle (HE) errors which indicate the lateral and angular distance respectively of the nose of the plane from the center-line of the runway. We use a model called the Tiny Taxinet~\cite{TinyTaxiNet} which takes in a down-sampled 8$\times$16 image of the runway. It has a training set and test set with 51462 and 7386 inputs respectively.

The tool, neural network models along with the datasets and Appendix are publicly available at the GitHub repository\footnote{\textcolor{blue}{\url{https://github.com/DNNCov/DNNCov}}}. All experiments were run on a machine with an Intel Core i9-9980HK processor and 64GB RAM running Windows 10. All experiments were repeated 5 times and average results are reported. 

\begin{table*}
	\centering
	\caption{Coverage of the Neural Network Models for the standard Test Sets.}
\label{tab:comparison}
\scalebox{0.8}{
\begin{tabular}{L{3.9cm}R{1.5cm}R{1.5cm}R{1.5cm}R{1.5cm}R{2.3cm}}
		 \hline
		   	    \textbf{Coverage Metrics (\%)}  &  \textbf{LeNet-1} & \textbf{LeNet-4}  & \textbf{LeNet-5} & \textbf{ResNet20}  & \textbf{TinyTaxiNet} \\
		   	    \hline
KMNC (K: 10) &95.0&85.29&90.7&98.75&62.35\\
KMNC (K: 1000) &60.23&54.33&59.10&71.16&43.54\\ \hline
TKNC (K: 10)&88.57&81.59&82.40&65.09&52.06\\
TKNC (K: 1000)&1.0&3.27&4.93&3.90&0.59\\ \hline

NBC         &0.87&0.66&0.58&5.55&2.01 \\ \hline
SNAC        &0.87&1.05&1.16&6.46&3.21\\ \hline
NC (Threshold: 0.00)&100.0&90.58&96.12&100.00&67.65\\ 
NC (Threshold: 0.20)&61.9 &76.09 &85.66 &100.00 &67.65\\
NC (Threshold: 0.50)&30.95 &63.77 &74.03 &99.16 &61.76\\
NC (Threshold: 0.75)&23.81&59.42&67.05&13.99&52.94\\
\hline
MC/DC (Sign-Sign)   & 5.77  & 44.17 & 10.62  &    24.66&27.60\\
MC/DC (Sign-Value)   & 63.46  &35.68  & 8.40  &    51.58&28.65\\
MC/DC (Value-Sign)   & 23.08  & 58.20 & 13.71  &   52.20&41.67\\
MC/DC (Value-Value)   & 100.00  &67.21  & 15.23  & 99.63   &46.88\\

               \hline

\end{tabular}
}
\end{table*}

\subsection{Comparison of coverage metrics} 
Table~\ref{tab:comparison} depicts the results obtained for different coverage metrics, in terms of percentage of covered obligations, when testing the models using the default test data from MNIST, CIFAR-10 and Tiny TaxiNet. The respective training sets were used to obtain threshold values for measures such as \textit{KMNC},  \textit{NBC}, \textit{SNAC} and \textit{MC/DC} metrics. As expected, \textit{NC (Threshold: 0.00)} seems the easiest to achieve. The \textit{NC} value is greater than 67\% for all the models, with 100\% for the LeNet-1 and even the most complex ResNet20 model. Setting the threshold to 0.75 reduces the values for all of the models. 

The \textit{KMNC (K: 10)} assesses if the tests cover different ranges of neuron values. We can observe that the value for this metric is greater than 85\% for the LeNet and ResNet20 models whereas it is only 62\% for Tiny TaxiNet. This appears reasonable considering that the test sets for LeNet and ResNet20 are the standard test sets for the widely studied MNIST and CIFAR-10 respectively, which are expected to test the respective models adequately. On the other hand, the test set for Tiny TaxiNet is generated by simulation which may not have a good coverage of different neuron values. A similar trend can be observed for \textit{TKNC (K: 10)}. 

The \textit{NBC} and \textit{SNAC} metrics determine if there are any tests that exercise neurons beyond the boundaries observed for the training sets. Looking at the results we can understand that the test set for the MNIST, CIFAR and Tiny TaxiNet benchmarks are very close in their distribution to the respective training sets which results in low values for these two metrics. 

The \textit{MC/DC} coverage appears to be more difficult to achieve than the other structural metrics. This is expected considering that it involves satisfaction of more constraints. 
The Value-Value variant has generally the highest coverage values and seems easier to achieve than the Sign-Sign counterparts. The test set for MNIST has better \textit{MC/DC (VV)} coverage values for LeNet-1 and LeNet-4, whereas the LeNet-5 model with a more complex architecture has the least \textit{MC/DC (VV)} coverage values. 

\begin{figure*}
\centering
\includegraphics[scale=0.7]{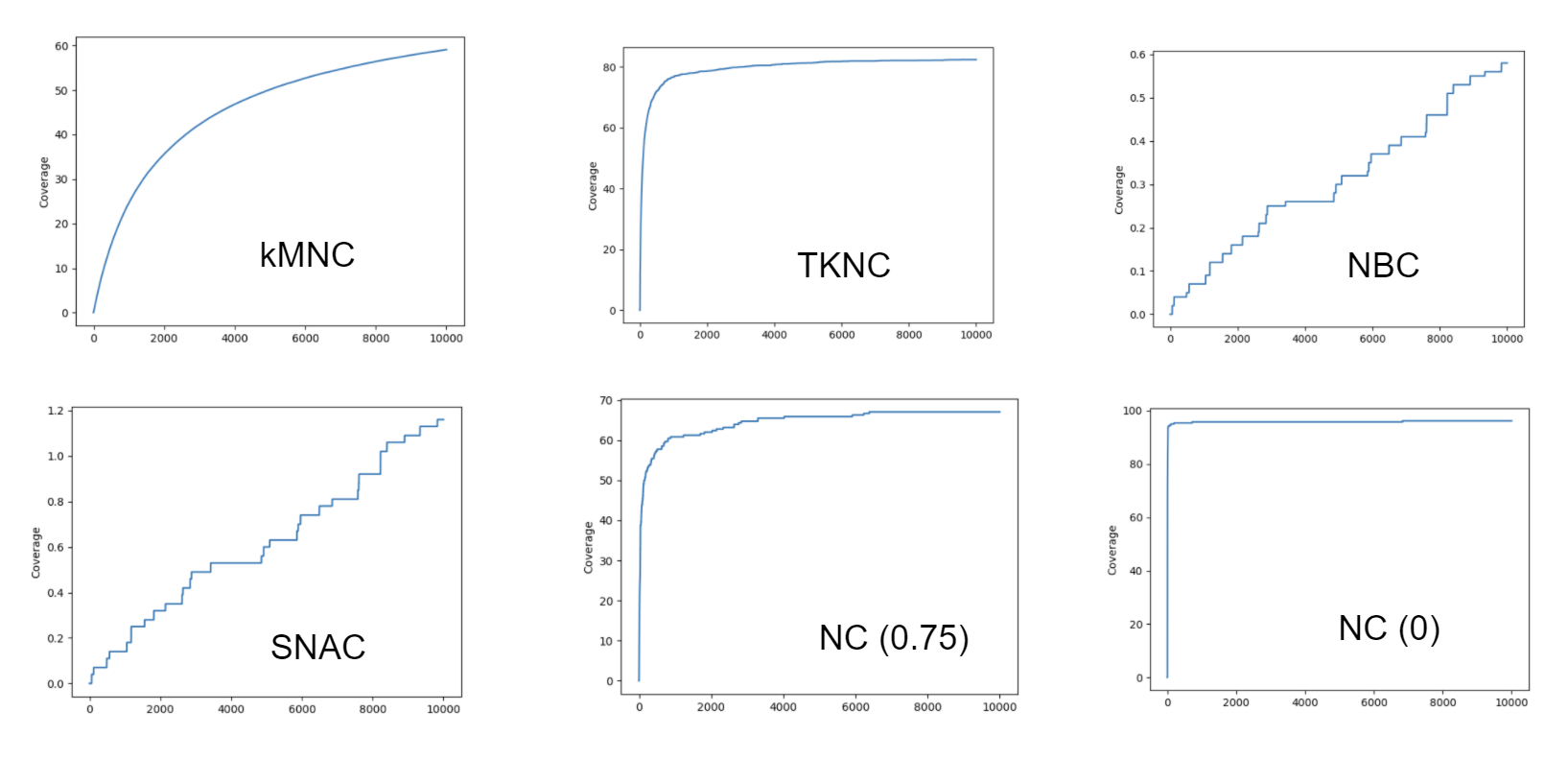}
\caption{Comparison of Coverage Metrics for LeNet-5.}
\label{fig:comparison}
\end{figure*}

Figure~\ref{fig:comparison} shows in more detail how the metrics compare for the same test set on the same model (LeNet-5). The graphs show how the coverage values change as the number of tests increase. As can be seen the \textit{KMNC}, \textit{NBC} and \textit{SNAC} metrics display a more gradual increase in comparison with \textit{NC} and \textit{TKNC}. This indicates that the \textit{KMNC}, \textit{NBC} and \textit{SNAC} may be better for indicating the quality of the test set better than the other metrics. For instance, the maximum values for \textit{TKNC} and \textit{NC} can be achieved by executing just 10\% of the test set. The additional tests which are redundant with respect to these metrics are actually useful since they cover different neuron value ranges. 

Also note that accuracy on the test sets is typically used to judge the quality of a trained model. Table~\ref{tab:subjects} provides the statistical test set accuracy of the models. 
For instance, LeNet-5 has 92.8\% accuracy while Tiny TaxiNet has Mean Absolute Error for ``CTE": 1.44 and for ``HE": 2.75. A user may consider these models to have good generalization and accuracy on unseen inputs. However, the corresponding coverage measures highlight how much this test set accuracy could be trusted. For instance, the Tiny TaxiNet model has low coverage scores indicating that test set may not have sufficiently tested the behaviors of the model.   

\subsection{Evaluating coverage metrics with respect to functional diversity} We use each DNN output class as a proxy for the functionality of the model. A dataset is considered to be functionally diverse if it contains inputs belonging to a comprehensive range of output classes. For example, a dataset which consists of inputs belonging to ten output classes has higher functional diversity than a dataset which consists of inputs belonging to only one output class (even if the structural coverage is the same). Intuitively, a functionally diverse test set is better in testing and debugging neural network models, as it covers more behaviors. If a test coverage metric is sensitive to functional diversity, it should obtain higher coverage when the dataset contains inputs belonging to multiple output classes, as opposed to fewer classes.

\begin{figure}
    \begin{subfigure}[b]{0.5\columnwidth}
        \includegraphics[width=\columnwidth]{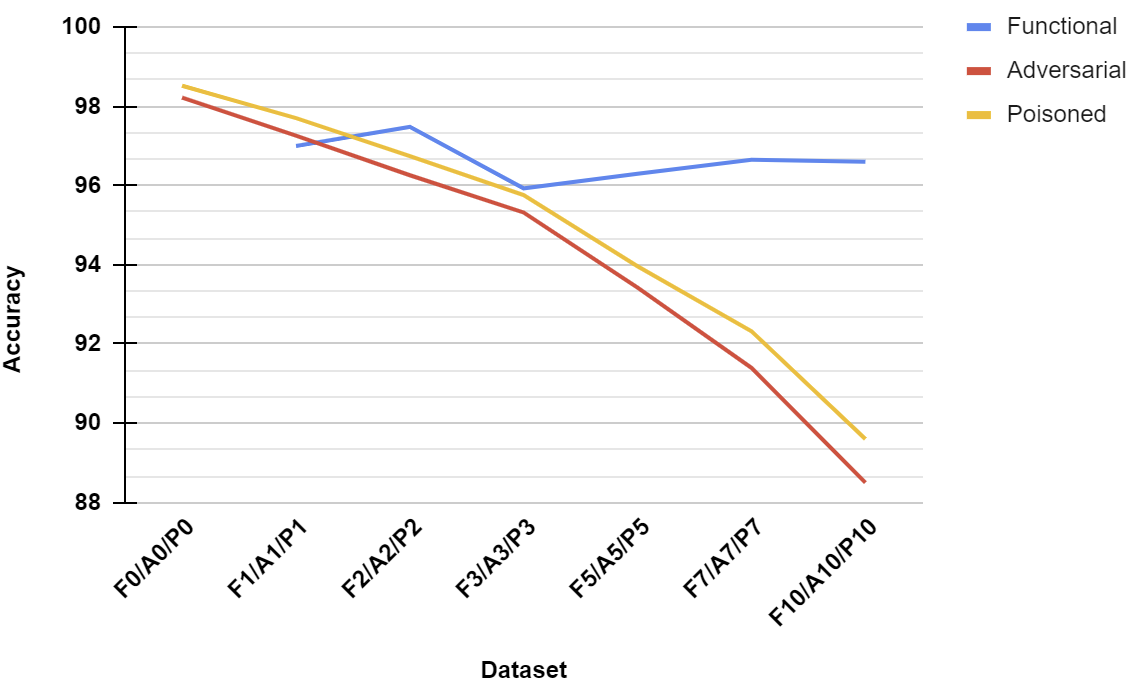}
        \caption{Accuracy}
        \label{fig:accuracy}
    \end{subfigure}
    \hfill
    \begin{subfigure}[b]{0.5\columnwidth}
        \includegraphics[width=\columnwidth]{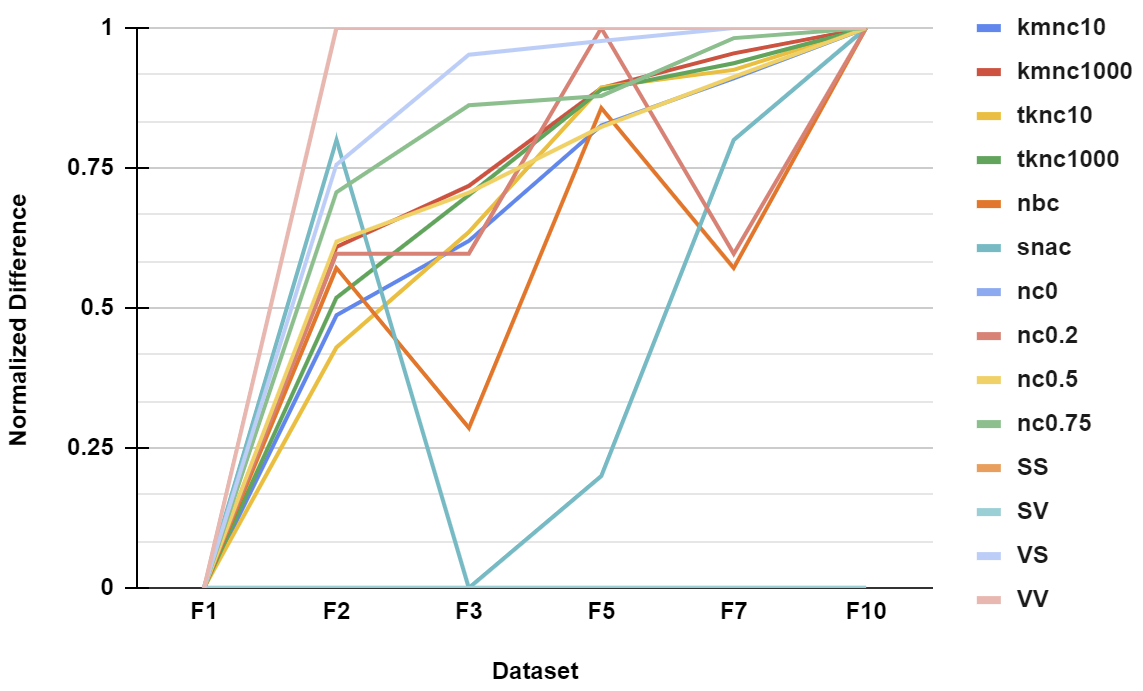}
        \caption{Functional Diversity}
        \label{fig:functional}
    \end{subfigure}
    \begin{subfigure}[b]{0.5\columnwidth}
        \includegraphics[width=\columnwidth]{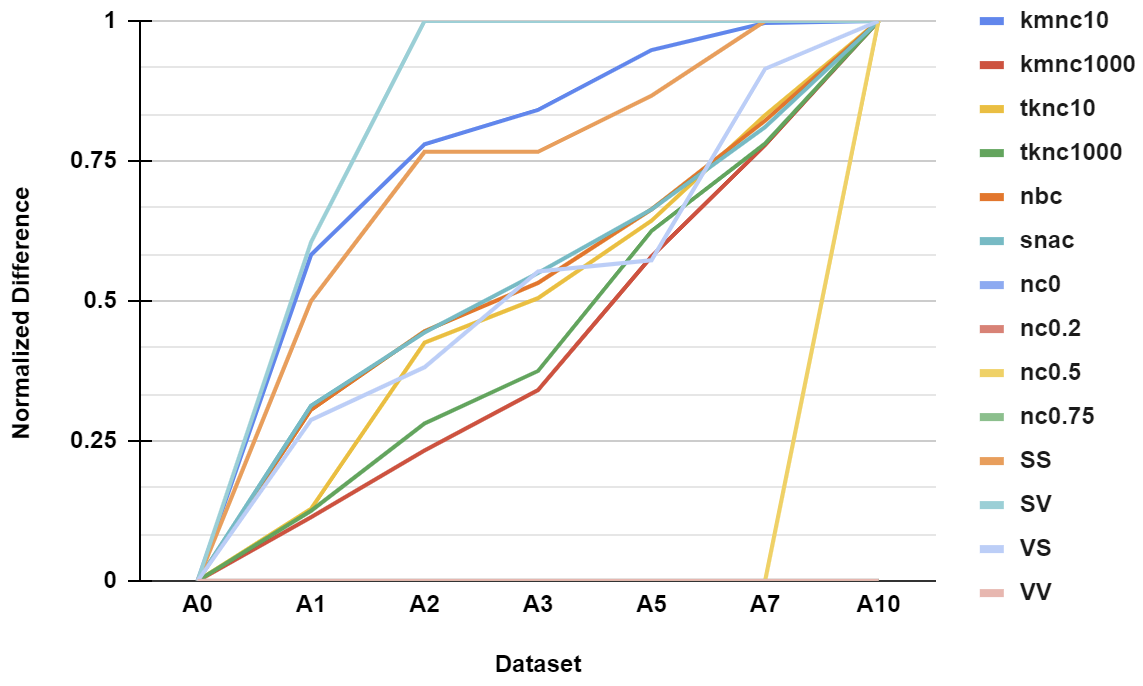}
        \caption{Defect Detection (Adversarial$_{\epsilon=0.30}$)}
        \label{fig:adversarial}
    \end{subfigure}
    \hfill
    \begin{subfigure}[b]{0.5\columnwidth}
        \includegraphics[width=\columnwidth]{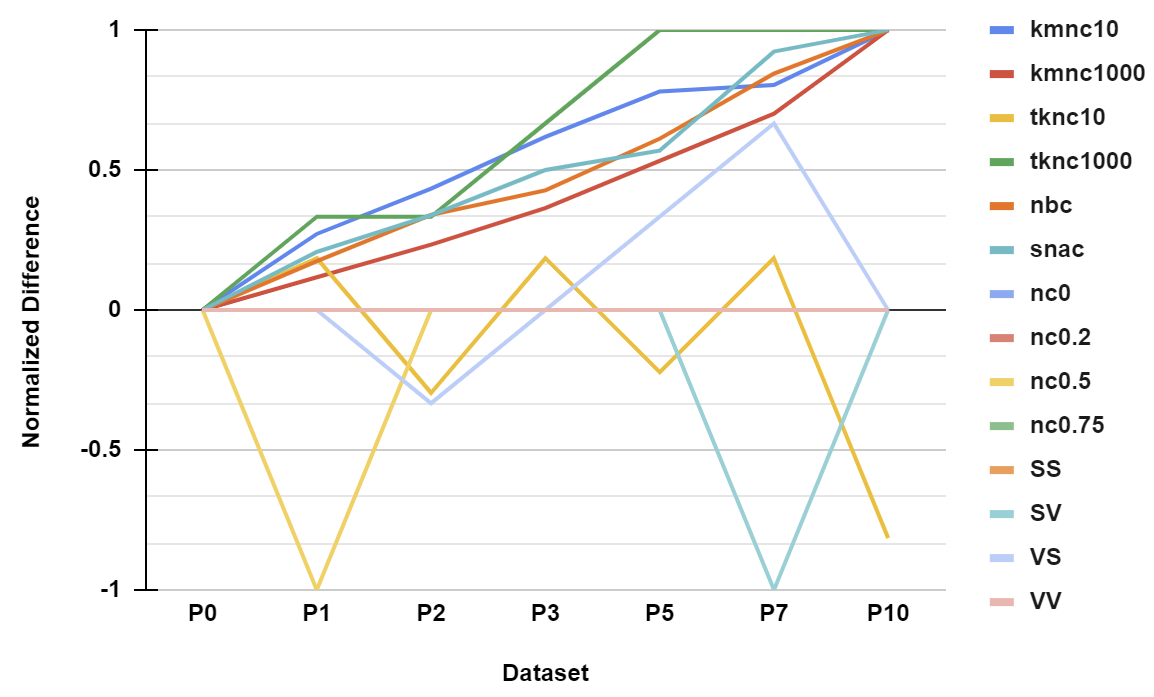}
        \caption{Defect Detection (Poisoned)}
        \label{fig:poisoned}
    \end{subfigure}
    \caption{Summary of Results}
    \label{fig:four figures}
\end{figure}

To evaluate the sensitivity of each metric towards functional diversity, we create six different datasets from the clean MNIST test set (\emph{MNIST-Test}). We name the datasets as F$_{\alpha}$ where $\alpha$ is the number of randomly selected output classes whose inputs are included in the respective dataset. For example, F$_{1}$ represents the dataset which consists of inputs belonging to one randomly selected output class and F$_{10}$ is the dataset which consists of inputs belonging to ten output classes. Similarly, we have F$_{2}$, F$_{3}$, F$_{5}$ and F$_{7}$. The size of each dataset is fixed to 800. 

To better compare the different coverage metrics, we normalize the results as follows.
For each dataset, we first calculate $\Delta_{\alpha}$ which is essentially the difference (Eq. \ref{eq:func-difference}) between the coverage value of the metric on the baseline dataset (F$_{1}$) and the respective dataset (F$_{\alpha}$). Max($\Delta_{\alpha}$) is the maximum $\Delta_{\alpha}$ across datasets F$_{1}$ to F$_{10}$. Min($\Delta_{\alpha}$) is the minimum $\Delta_{\alpha}$ across datasets F$_{1}$ to F$_{10}$. We then calculate normalized difference (NCoverage(F$_{\alpha}$)) in coverage (Eq. \ref{eq:func-normalized}). 
\begin{equation*}
\tag{1}
\label{eq:func-difference}
\Delta_{\alpha}=Coverage_{F_{\alpha}}-Coverage_{F_{1}}
\end{equation*}
\begin{equation*}
\tag{2}
\label{eq:func-normalized}
NCoverage(F_{\alpha}) = \frac{Coverage_{F_{\alpha}}-Coverage_{F_{1}}}{Max(\Delta_{\alpha})-Min(\Delta_{\alpha})}
\end{equation*}


Figure \ref{fig:functional} summarizes the results. Detailed results are available in the Appendix. We consider the coverage obtained on dataset $F_{1}$ as the baseline. We can see that as the value of $\alpha$ is increased from 1 to 10, the metrics report higher coverage (shown by the high normalized difference value). With the exception of \textit{NC (Threshold: 0.00)}, \textit{MC/DC (SS)} and \textit{MC/DC (SV)}, all of the remaining 11 metrics/configurations shows highest coverage on dataset F$_{10}$. \textit{NC (Threshold: 0.00)} achieves 100\% coverage on all datasets (even when the dataset consists of inputs belonging to only one output class) whereas \textit{MCDC(SS)} and \textit{MC/DC (SV)} obtains 0\% coverage on all datasets (even when the dataset has inputs belonging to all output classes). Due to this issue, we cannot observe any difference in these three coverage metrics across all the datasets. 

Figure \ref{fig:accuracy} and Table \ref{tab:accuracy} shows that the accuracy of the neural network model remains between 95\% and 98\%. Thus, the accuracy metric fails to differentiate between these datasets.

\begin{table*}
	\centering
	\caption{Accuracy of Model on Datasets. Column "Dataset" shows the name of the Datasets. Column "MNIST (Functional)" shows the Accuracy for Datasets F$_{1}$ to F$_{10}$. Column "MNIST (Adversarial)" shows the Accuracy for Datasets A$_{0}$ to A$_{10}$. Column "MNIST (Poisoned)" shows the Accuracy for Datasets P$_{0}$ to P$_{10}$.}
\label{tab:accuracy}
\begin{tabular}{L{2.0cm}C{3.0cm}C{3.0cm}C{2.3cm}}
		 \hline
		   	    \multirow{2}{*}{\textbf{Dataset}}  &  \textbf{MNIST (Functional)} & \textbf{MNIST (Adversarial)}  & \textbf{MNIST (Poisoned)} \\
		   	    \hline

F$_{0}$/A$_{0}$/P$_{0}$  &N/A  &98.22  &98.52      \\\hline
F$_{1}$/A$_{1}$/P$_{1}$  &97.00   &	97.26 &   97.70	    \\\hline
F$_{2}$/A$_{2}$/P$_{2}$  &97.48   &	96.26 &   96.74        	\\\hline
F$_{3}$/A$_{3}$/P$_{3}$  &95.93   &	95.32 &   95.76        	\\\hline
F$_{5}$/A$_{5}$/P$_{5}$  &96.30   &	93.42 &   93.96        	\\\hline
F$_{7}$/A$_{7}$/P$_{7}$  & 96.65    &	91.40&92.32           	\\\hline
F$_{10}$/A$_{10}$/P$_{10}$ & 96.60  &	88.50  &89.60           	\\\hline
\end{tabular}
\end{table*}


\subsection{Evaluating coverage metrics with respect to defect detection.} We evaluate the different metrics by testing the  defect detection ability, with respect to adversarial robustness and data poisoning scenarios.
\subsubsection{Adversarial Scenario.}
In an adversarial attack \cite{szegedy2013intriguing}, a neural network mis-classifies the adversarial input that is obtained by adding adversarial perturbations on the original input. 
Testing the neural network model with adversarial inputs can help one identify if the neural network is robust towards adversarial attacks. 

\label{sec:defect}

For this scenario, we apply perturbations to MNIST images in the training (\emph{MNIST-Train}) and test sets (\emph{MNIST-Test}) using the FGSM (Fast Gradient Sign Method) attack \cite{szegedy2013intriguing}, with $\epsilon\in$[0.01, 0.05, 0.10, 0.20, 0.30] being used for controling the perturbation level. This gives us \emph{MNIST-Adversarial-Train} and \emph{MNIST-Adversarial-Test} respectively. Due to space limitations, we report results for $\epsilon$ = 0.30 in this paper. Results for other values of $\epsilon$ are available in the Appendix. Intuitively, if a coverage metric has the capability to check the defect detection ability of a dataset, it should obtain higher coverage when the dataset contains inputs that can reveal defects in the neural network model. We create dataset A$_{0}$ by randomly selecting 1000 inputs from the clean MNIST test set (\emph{MNIST-Test}). We consider the coverage obtained on this dataset as the baseline. From this baseline dataset (A$_{0}$), we further create six different datasets by replacing $\beta$\% of clean inputs with adversarial inputs. We name these datasets as A$_{\beta}$. For example, A$_{1}$ represents the dataset in which we replaced 1\% (10) of the clean inputs with the same number of randomly selected adversarial inputs and A$_{10}$ is the dataset in which we replaced 10\% (100) of the clean inputs with the same number of randomly selected adversarial inputs. Similarly, we have A$_{2}$, A$_{3}$, A$_{5}$ and A$_{7}$. The size of each dataset is 1000. 

For each dataset, we first calculate $\Delta_{\beta}$ which is essentially the difference (Eq. \ref{eq:adv-difference}) between the coverage value of the metric on the baseline dataset (A$_{0}$) and the respective dataset (A$_{\beta}$). Max($\Delta_{\beta}$) is the maximum $\Delta_{\beta}$ across datasets A$_{0}$ to A$_{10}$. Min($\Delta_{\beta}$) is the minimum $\Delta_{\beta}$ across datasets A$_{0}$ to A$_{10}$.
We then calculate normalized difference (NCoverage(A$_{\beta}$)) in coverage (Eq. \ref{eq:adv-normalized}). 
\begin{equation*}
\tag{3}
\label{eq:adv-difference}
\Delta_{\beta}=Coverage_{A_{\beta}}-Coverage_{A_{0}}
\end{equation*}
\begin{equation*}
\tag{4}
\label{eq:adv-normalized}
NCoverage(A_{\beta}) = \frac{Coverage_{A_{\beta}}-Coverage_{A_{0}}}{Max(\Delta_{\beta})-Min(\Delta_{\beta})}
\end{equation*}

Figure \ref{fig:adversarial} summarizes the results. Detailed results are available in the Appendix. We can see that as the value of $\beta$ is increased from 0 to 10, ten metrics (with the exception of \textit{NC (Threshold: 0.00)}, \textit{NC (Threshold:  0.20)}, \textit{NC (Threshold: 0.75)} and \textit{MC/DC (VV)}) report higher coverage (shown by the high normalized difference value). \textit{NC (Threshold: 0.00)}, \textit{NC (Threshold: 0.20)}, \textit{NC (Threshold: 0.75)} and \textit{MC/DC (VV)} report similar coverage across all datasets. Due to this issue, we cannot observe any difference in these coverage metrics across all the datasets. 
Figure \ref{fig:accuracy} and Table \ref{tab:accuracy} shows that the accuracy of the model is highest on clean dataset (A$_{0}$) and it gradually decreases as the value of $\beta$ is increased. This is expected because as the number of adversarial inputs increases in the dataset, there are more inputs on which the neural network fails to predict the correct class. 


\subsubsection{Poisoned Scenario.}
In a data poisoning scenario~\cite{badnets}, an attacker has 'poisoned' the training data such that the model has high accuracy on clean data but when it is presented with an input that contains a poisoned 'trigger', the model mis-classifies the respective input to a target output class. Testing the neural network model with poisoned inputs can help one identify if the neural network contains a backdoor which can be exploited by a malicious actor. 

\begin{figure}
\centering
  \includegraphics[]{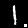}
  \includegraphics[]{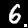}
  \includegraphics[]{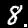}
  \hspace{0.3cm}
  \includegraphics[]{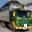}
  \includegraphics[]{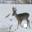}
  \includegraphics[]{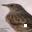}
  \caption{Example poisoned data for MNIST (left) and CIFAR-10 (right). The backdoor trigger is embedded as the white square at the bottom right corner of each image. When the backdoor appears, the poisoned MNIST model will classify the input as {\lab 7} and the poisoned
  CIFAR-10 model will classify it as {\lab {horse}}.}
  \label{fig:poisonedexample}
  \vspace{-5mm}
\end{figure}

For the poisoned scenario, we apply the backdoor attack from \cite{badnets}. Figure \ref{fig:poisonedexample}
shows some examples of poisoned attacks. We applied the backdoor trigger to the first 600 inputs in the MNIST training set (\emph{MNIST-Train}); rest of the 59400 inputs are not changed. This results in the \emph{MNIST-Poisoned-Train} set. For evaluation, we created a poisoned test set (\emph{MNIST-Poisoned-Test}) from the MNIST test set (\emph{MNIST-Test}).

We create dataset P$_{0}$ by randomly selecting 1000 inputs from the clean MNIST test set (\emph{MNIST-Test}). We consider the coverage obtained on this dataset as the baseline. From this baseline dataset (P$_{0}$), we further create six different datasets by replacing $\beta$\% of clean inputs with poisoned inputs. We name these datasets as P$_{\beta}$. For example, P$_{1}$ represents the dataset in which we replaced 1\% (10) of the clean inputs with the same number of randomly selected poisoned inputs and P$_{10}$ is the dataset in which we replaced 10\% (100) of the clean inputs with the same number of randomly selected poisoned inputs. Similarly, we have P$_{2}$, P$_{3}$, P$_{5}$ and P$_{7}$. The size of each dataset is 1000. 

For each dataset, we first calculate $\Delta_{\beta}$ which is essentially the difference (similar to Eq. \ref{eq:adv-difference}) between the coverage value of the metric on the baseline dataset (P$_{0}$) and the respective dataset (P$_{\beta}$). Max($\Delta_{\beta}$) is the maximum $\Delta_{\beta}$ across datasets P$_{0}$ to P$_{10}$. Min($\Delta_{\beta}$) is the minimum $\Delta_{\beta}$ across datasets P$_{0}$ to P$_{10}$.
We then calculate normalized difference (NCoverage(P$_{\beta}$)) in coverage (similar to Eq. \ref{eq:adv-normalized}).

We repeated similar experiments using poisoned data. Figure \ref{fig:poisoned} summarizes the results. Detailed results are available in the Appendix. We can see that as the value of $\beta$ is increased from 0 to 10, five metrics (\textit{KMNC (K: 10)}, \textit{KMNC (K: 1000)}, \textit{TKNC (K: 1000)}, \textit{NBC} and \textit{SNAC}) report higher coverage (shown by the high normalized difference value). All variants of \textit{NC} and \textit{MC/DC} fail to differentiate between the datasets because they report almost similar coverage across all datasets. Four of the metrics (\textit{TKNC (K: 10)}, \textit{NC (Threshold: 0.50)}, \textit{MC/DC (SV)} and \textit{MC/DC (VS)}) even report lesser coverage on poisoned datasets.  \textit{NC (Threshold: 0.00)} and \textit{MC/DC (VV)} achieves 100\% coverage on all datasets (even when the dataset does not contain any poisoned inputs) and \textit{MC/DC (SS)} obtain 0\% coverage on all datasets (even when the dataset consists 10\% of poisoned inputs). Due to this, we cannot observe any difference in these coverage metrics across all the datasets. Therefore, these metrics are not good for checking the defect detection ability of the dataset. 


\begin{table*}[t]
	\centering
	\caption{Quantitative Information for Datasets.}
\label{tab:quantitative-summary}
\scalebox{0.35}{\begin{tabular}{C{2.8cm}C{1.3cm}C{1.0cm}R{1.2cm}R{1.3cm}R{1.3cm}R{1.2cm}R{0.8cm}R{1.0cm}R{1.3cm}R{1.4cm}R{1.5cm}R{1.6cm}R{2.2cm}R{2.4cm}R{2.4cm}R{3.0cm}}
		 \hline
		   	    \multirow{2}{*}{\textbf{Scenario}}& \multirow{2}{*}{\textbf{Dataset}}  &  \multirow{2}{*}{\textbf{Metric}}&  \textbf{KMNC (10)} & \textbf{KMNC (1000)} &\textbf{TKNC (10)}&\textbf{TKNC (1000)}  &  \multirow{2}{*}{\textbf{NBC}}& \multirow{2}{*}{\textbf{SNAC}}  & \multirow{2}{*}{\textbf{NC (0)}} & \multirow{2}{*}{\textbf{NC (0.2)}} &\multirow{2}{*}{\textbf{NC (0.5)}} &\multirow{2}{*}{\textbf{NC (0.75)}}& \multirow{2}{*}{\textbf{MC/DC (SS)}}& \multirow{2}{*}{\textbf{MC/DC (SV)}}& \multirow{2}{*}{\textbf{MC/DC (VS)}}& \multirow{2}{*}{\textbf{MC/DC (VV)}}\\
		   	    \hline
\multirow{10}{*}{Functional} &\multirow{5}{*}{F$_{1}$} &Min&0.00&0.00&0.00&0.00&0.00&0.00&2.80&0.00&0.00&0.00&0.00&0.00&0.00&216.20\\
&&Max&254.60&16.00&245.60&245.60&0.80&0.80&251.00&251.40&253.40&250.40&0.00&0.00&47,428.40&80,757.40\\
&&Avg&44.26&0.80&9.96&0.74&0.00&0.00&40.91&64.62&85.99&42.92&0.00&0.00&14,994.11&15,958.12\\
&&Std.&67.43&1.67&28.09&4.80&0.02&0.03&41.07&67.64&82.06&63.53&0.00&0.00&14,461.88&18,205.57\\
&&Var.&4,546.93&2.81&789.82&23.07&0.00&0.00&1,701.20&4,622.98&6,738.52&4,056.32&0.00&0.00&217,047,915.68&343,441,740.25\\\cline{2-17}
&\multirow{5}{*}{F$_{10}$}&Min&0.00&0.00&0.00&0.00&0.00&0.00&3.20&0.00&0.00&0.00&0.00&0.00&0.00&7,972.00\\
&&Max&254.80&10.80&228.80&228.80&0.80&0.80&246.80&254.40&253.80&252.60&0.00&0.00&90,890.80&142,748.00\\
&&Avg&72.01&0.80&11.23&0.75&0.00&0.00&39.53&80.42&110.70&80.85&0.00&0.00&44,167.25&40,287.12\\
&&Std.&76.94&1.25&24.34&4.01&0.02&0.03&39.64&83.46&81.18&75.04&0.00&0.00&28,187.31&32,664.23\\
&&Var.&5,919.58&1.55&592.63&16.11&0.00&0.00&1,573.37&6,967.05&6,590.26&5,630.99&0.00&0.00&795,678,292.52&1,069,718,510.41\\\hline 

\multirow{10}{*}{Adversarial$_{\epsilon=0.30}$} &\multirow{5}{*}{A$_{0}$} &Min&0.00&0.00&0.00&0.00&0.00&0.00&0.20&0.00&0.00&0.00&0.00&0.00&0.00&0.00\\
&&Max&255.00&12.60&243.80&246.40&0.80&0.80&232.00&246.00&252.20&244.80&0.00&98,840.40&125,897.80&199,146.00\\
&&Avg&74.26&0.99&14.98&0.94&0.00&0.00&218.81&196.66&94.57&55.21&0.00&749.13&20,325.60&35,183.88\\
&&Std.&78.73&1.42&33.36&6.42&0.02&0.03&40.92&61.28&73.75&71.70&0.00&7,946.83&40,408.32&44,938.10\\
&&Var.&6,198.64&2.03&1,112.89&41.19&0.00&0.00&1,674.34&3,755.88&5,440.95&5,143.82&0.00&63,197,042.79&1,633,247,596.37&2,019,902,359.70\\\cline{2-17}
&\multirow{5}{*}{A$_{10}$}&Min&0.00&0.00&0.00&0.00&0.00&0.00&0.00&0.00&0.00&0.00&0.00&0.00&0.00&0.00\\
&&Max&255.00&11.60&233.00&247.20&69.20&69.20&232.00&251.60&253.40&246.60&2,363.60&95,831.60&146,695.20&242,320.40\\
&&Avg&80.15&0.98&14.91&0.94&0.34&0.68&218.81&194.41&93.05&56.35&6.08&797.94&22,522.68&44,255.64\\
&&Std.&78.65&1.37&32.04&6.25&3.17&4.46&39.96&60.96&73.37&73.38&114.16&7,496.02&44,955.50&58,645.70\\
&&Var.&6,184.81&1.88&1,026.34&39.04&10.10&19.96&1,596.77&3,717.62&5,384.96&5,386.06&13,825.67&56,250,911.82&2,021,423,808.66&3,439,790,761.82\\\hline 

\multirow{10}{*}{Poisoned} & \multirow{5}{*}{P$_{0}$} &Min&0.00&0.00&0.00&0.00&0.00&0.00&30.80&0.00&0.00&0.00&0.00&0.00&0.00&12,986.40\\
&&Max&255.00&12.60&247.00&250.20&1.00&1.00&232.60&247.40&253.80&248.20&0.00&156,211.40&166,161.00&219,296.00\\
&&Avg&65.99&1.00&15.84&0.95&0.00&0.00&221.60&199.85&122.49&99.68&0.00&882.18&66,915.34&43,847.31\\
&&Std.&74.18&1.49&33.46&5.05&0.03&0.03&32.65&56.13&77.95&73.52&0.00&10,372.68&38,601.23&31,083.70\\
&&Var.&5,503.32&2.21&1,120.00&25.46&0.00&0.00&1,071.57&3,151.56&6,076.93&5,405.70&0.00&107,599,179.64&1,490,070,791.53&966,466,131.37\\\cline{2-17}
&\multirow{5}{*}{P$_{10}$}&Min&0.00&0.00&0.00&0.00&0.00&0.00&35.40&0.00&0.00&0.00&0.00&0.00&0.00&13,404.20\\
&&Max&255.00&12.00&247.40&247.40&64.60&64.60&234.20&246.80&254.40&249.80&0.00&155,748.20&162,128.00&217,657.80\\
&&Avg&68.33&1.00&15.74&0.95&0.03&0.07&221.95&199.37&118.89&95.75&0.00&878.16&67,644.00&45,336.55\\
&&Std.&75.30&1.46&32.92&4.98&1.15&1.63&31.16&54.54&78.35&73.83&0.00&10,327.47&38,962.82&30,707.17\\
&&Var.&5,669.85&2.15&1,084.01&24.84&1.33&2.66&971.28&2,975.74&6,139.18&5,452.29&0.00&106,676,969.06&1,518,177,548.27&943,247,524.91\\\hline
\end{tabular}
}
\end{table*}

\subsection{Quantitative and Visualization information.} We also discuss briefly the quantitative information that can be computed with \dnncov. We calculate minimum, maximum, average, standard deviation and variance of the number of inputs that cover each of the coverage obligations for each coverage criterion. The summarized results are shown in Table \ref{tab:quantitative-summary}. Detailed results are available in the Appendix. 
Different test suites may have similar total coverage values but different coverage distributions. This cannot be highlighted by existing coverage metrics. For example, in Section \ref{sec:defect}, we mentioned that \textit{MC/DC (VV)} remained constant across all datasets (A$_{0}$ to A$_{10}$). However, if we analyze the quantitative information (see A$_{0}$ and A$_{10}$ rows in Table \ref{tab:quantitative-summary}), we can see that even though the \emph{MC/DC (VV)} remained 100\% across A$_{0}$ and A$_{10}$, on average more input pairs (44255 vs 35183) covered each neuron pair on A$_{10}$ as compared to A$_{0}$. This shows that dataset A$_{10}$ is better than dataset A$_{0}$ and it may reveal more defects in the neural network model. 

Figure \ref{fig:visualization} shows the visualisation report generated by \dnncov. We can see that neurons in layer 5 are covered fewer times as compared to neurons in the layer 1. Moreover, we can observe that neuron 4 in layer 2 is never covered. Either this neuron is not required and it can be pruned from the neural network or the test set has not properly tested this neuron. 

\section{Conclusion}
\label{sec:conclusion}
In this paper we presented and evaluated several recently proposed structural coverage criteria for testing neural networks. We also described \dnncov, an integrated tool  for helping developers with testing neural networks, by providing the means to measure, visualize and compare different, state-of-the-art testing criteria for neural networks. The results indicate that many of the existing structural metrics are not sensitive to functional diversity and defect detection abilities in test suites. To address these limitations, we plan to develop new coverage metrics that are still in terms of the structure of the DNN, but also have  semantic meaning, such as activation patterns \cite{DBLP:conf/kbse/GopinathCPT19}. We also plan to add more structural coverage criteria to the tool, as they appear in the related research.
We are also working on a method for measuring KMNC, NBC, and SNAC coverage in the absence of the training set. One idea is to classify statistical outliers as {\em boundary} cases that would be part of NBC/SNAC and determine KMNC based on the non-boundary cases. Finally, we plan to improve the visualization component in \dnncov to display more information about the network architecture.

\clearpage
\bibliography{all.bib}


\begin{thebibliography}{18}
\ifx \bisbn   \undefined \def \bisbn  #1{ISBN #1}\fi
\ifx \binits  \undefined \def \binits#1{#1}\fi
\ifx \bauthor  \undefined \def \bauthor#1{#1}\fi
\ifx \batitle  \undefined \def \batitle#1{#1}\fi
\ifx \bjtitle  \undefined \def \bjtitle#1{#1}\fi
\ifx \bvolume  \undefined \def \bvolume#1{\textbf{#1}}\fi
\ifx \byear  \undefined \def \byear#1{#1}\fi
\ifx \bissue  \undefined \def \bissue#1{#1}\fi
\ifx \bfpage  \undefined \def \bfpage#1{#1}\fi
\ifx \blpage  \undefined \def \blpage #1{#1}\fi
\ifx \burl  \undefined \def \burl#1{\textsf{#1}}\fi
\ifx \doiurl  \undefined \def \doiurl#1{\url{https://doi.org/#1}}\fi
\ifx \betal  \undefined \def \betal{\textit{et al.}}\fi
\ifx \binstitute  \undefined \def \binstitute#1{#1}\fi
\ifx \binstitutionaled  \undefined \def \binstitutionaled#1{#1}\fi
\ifx \bctitle  \undefined \def \bctitle#1{#1}\fi
\ifx \beditor  \undefined \def \beditor#1{#1}\fi
\ifx \bpublisher  \undefined \def \bpublisher#1{#1}\fi
\ifx \bbtitle  \undefined \def \bbtitle#1{#1}\fi
\ifx \bedition  \undefined \def \bedition#1{#1}\fi
\ifx \bseriesno  \undefined \def \bseriesno#1{#1}\fi
\ifx \blocation  \undefined \def \blocation#1{#1}\fi
\ifx \bsertitle  \undefined \def \bsertitle#1{#1}\fi
\ifx \bsnm \undefined \def \bsnm#1{#1}\fi
\ifx \bsuffix \undefined \def \bsuffix#1{#1}\fi
\ifx \bparticle \undefined \def \bparticle#1{#1}\fi
\ifx \barticle \undefined \def \barticle#1{#1}\fi
\bibcommenthead
\ifx \bconfdate \undefined \def \bconfdate #1{#1}\fi
\ifx \botherref \undefined \def \botherref #1{#1}\fi
\ifx \url \undefined \def \url#1{\textsf{#1}}\fi
\ifx \bchapter \undefined \def \bchapter#1{#1}\fi
\ifx \bbook \undefined \def \bbook#1{#1}\fi
\ifx \bcomment \undefined \def \bcomment#1{#1}\fi
\ifx \oauthor \undefined \def \oauthor#1{#1}\fi
\ifx \citeauthoryear \undefined \def \citeauthoryear#1{#1}\fi
\ifx \endbibitem  \undefined \def \endbibitem {}\fi
\ifx \bconflocation  \undefined \def \bconflocation#1{#1}\fi
\ifx \arxivurl  \undefined \def \arxivurl#1{\textsf{#1}}\fi
\csname PreBibitemsHook\endcsname

\bibitem{pei2017deepxplore}
\begin{bchapter}
\bauthor{\bsnm{Pei}, \binits{K.}},
\bauthor{\bsnm{Cao}, \binits{Y.}},
\bauthor{\bsnm{Yang}, \binits{J.}},
\bauthor{\bsnm{Jana}, \binits{S.}}:
\bctitle{Deep{X}plore: Automated whitebox testing of deep learning systems}.
In: \bbtitle{SOSP}
(\byear{2017})
\end{bchapter}
\endbibitem

\bibitem{ma2018deepgauge}
\begin{bchapter}
\bauthor{\bsnm{Ma}, \binits{L.}},
\bauthor{\bsnm{Juefei-Xu}, \binits{F.}},
\bauthor{\bsnm{Zhang}, \binits{F.}},
\bauthor{\bsnm{Sun}, \binits{J.}},
\bauthor{\bsnm{Xue}, \binits{M.}},
\bauthor{\bsnm{Li}, \binits{B.}},
\bauthor{\bsnm{Chen}, \binits{C.}},
\bauthor{\bsnm{Su}, \binits{T.}},
\bauthor{\bsnm{Li}, \binits{L.}},
\bauthor{\bsnm{Liu}, \binits{Y.}}, \betal:
\bctitle{Deep{G}auge: Multi-granularity testing criteria for deep learning
  systems}.
In: \bbtitle{ASE}
(\byear{2018})
\end{bchapter}
\endbibitem

\bibitem{sun2019structural}
\begin{botherref}
\oauthor{\bsnm{Sun}, \binits{Y.}},
\oauthor{\bsnm{Huang}, \binits{X.}},
\oauthor{\bsnm{Kroening}, \binits{D.}},
\oauthor{\bsnm{Sharp}, \binits{J.}},
\oauthor{\bsnm{Hill}, \binits{M.}},
\oauthor{\bsnm{Ashmore}, \binits{R.}}:
Structural test coverage criteria for deep neural networks.
ACM Transactions on Embedded Computing Systems (TECS)
(2019)
\end{botherref}
\endbibitem

\bibitem{10.1145/3395363.3397346}
\begin{bchapter}
\bauthor{\bsnm{Lee}, \binits{S.}},
\bauthor{\bsnm{Cha}, \binits{S.}},
\bauthor{\bsnm{Lee}, \binits{D.}},
\bauthor{\bsnm{Oh}, \binits{H.}}:
\bctitle{Effective white-box testing of deep neural networks with adaptive
  neuron-selection strategy}.
In: \bbtitle{ISSTA}
(\byear{2020})
\end{bchapter}
\endbibitem

\bibitem{deephunter}
\begin{bchapter}
\bauthor{\bsnm{Xie}, \binits{X.}},
\bauthor{\bsnm{Ma}, \binits{L.}},
\bauthor{\bsnm{Juefei-Xu}, \binits{F.}},
\bauthor{\bsnm{Xue}, \binits{M.}},
\bauthor{\bsnm{Chen}, \binits{H.}},
\bauthor{\bsnm{Liu}, \binits{Y.}},
\bauthor{\bsnm{Zhao}, \binits{J.}},
\bauthor{\bsnm{Li}, \binits{B.}},
\bauthor{\bsnm{Yin}, \binits{J.}},
\bauthor{\bsnm{See}, \binits{S.}}:
\bctitle{Deep{H}unter: A coverage-guided fuzz testing framework for deep neural
  networks}.
In: \bbtitle{ISSTA}
(\byear{2019})
\end{bchapter}
\endbibitem

\bibitem{sun2018concolic}
\begin{bchapter}
\bauthor{\bsnm{Sun}, \binits{Y.}},
\bauthor{\bsnm{Wu}, \binits{M.}},
\bauthor{\bsnm{Ruan}, \binits{W.}},
\bauthor{\bsnm{Huang}, \binits{X.}},
\bauthor{\bsnm{Kwiatkowska}, \binits{M.}},
\bauthor{\bsnm{Kroening}, \binits{D.}}:
\bctitle{Concolic testing for deep neural networks}.
In: \bbtitle{ASE}
(\byear{2018})
\end{bchapter}
\endbibitem

\bibitem{EvalDNN}
\begin{bchapter}
\bauthor{\bsnm{Tian}, \binits{Y.}},
\bauthor{\bsnm{Zeng}, \binits{Z.}},
\bauthor{\bsnm{Wen}, \binits{M.}},
\bauthor{\bsnm{Liu}, \binits{Y.}},
\bauthor{\bsnm{Kuo}, \binits{T.-y.}},
\bauthor{\bsnm{Cheung}, \binits{S.-C.}}:
\bctitle{Evaldnn: A toolbox for evaluating deep neural network models}.
In: \bbtitle{2020 IEEE/ACM 42nd International Conference on Software
  Engineering: Companion Proceedings (ICSE-Companion)},
pp. \bfpage{45}--\blpage{48}
(\byear{2020})
\end{bchapter}
\endbibitem

\bibitem{huang2017safety}
\begin{bchapter}
\bauthor{\bsnm{Huang}, \binits{X.}},
\bauthor{\bsnm{Kwiatkowska}, \binits{M.}},
\bauthor{\bsnm{Wang}, \binits{S.}},
\bauthor{\bsnm{Wu}, \binits{M.}}:
\bctitle{Safety verification of deep neural networks}.
In: \bbtitle{CAV}
(\byear{2017})
\end{bchapter}
\endbibitem

\bibitem{kim2019guiding}
\begin{bchapter}
\bauthor{\bsnm{Kim}, \binits{J.}},
\bauthor{\bsnm{Feldt}, \binits{R.}},
\bauthor{\bsnm{Yoo}, \binits{S.}}:
\bctitle{Guiding deep learning system testing using surprise adequacy}.
In: \bbtitle{ICSE}
(\byear{2019})
\end{bchapter}
\endbibitem

\bibitem{MNIST}
\begin{botherref}
\oauthor{\bsnm{Deng}, \binits{L.}}:
The {MNIST} database of handwritten digit images for machine learning research.
IEEE Signal Processing Magazine
(2012)
\end{botherref}
\endbibitem

\bibitem{LENET1}
\begin{botherref}
\oauthor{\bsnm{LeCun}, \binits{Y.}},
\oauthor{\bsnm{Bottou}, \binits{L.}},
\oauthor{\bsnm{Bengio}, \binits{Y.}},
\oauthor{\bsnm{Haffner}, \binits{P.}}:
Gradient-based learning applied to document recognition.
Proceedings of the IEEE
(1998)
\end{botherref}
\endbibitem

\bibitem{LENET2}
\begin{bchapter}
\bauthor{\bsnm{LeCun}, \binits{Y.}},
\bauthor{\bsnm{Boser}, \binits{B.E.}},
\bauthor{\bsnm{Denker}, \binits{J.S.}},
\bauthor{\bsnm{Henderson}, \binits{D.}},
\bauthor{\bsnm{Howard}, \binits{R.E.}},
\bauthor{\bsnm{Hubbard}, \binits{W.E.}},
\bauthor{\bsnm{Jackel}, \binits{L.D.}}:
\bctitle{Handwritten digit recognition with a back-propagation network}.
In: \bbtitle{NIPS}
(\byear{1989})
\end{bchapter}
\endbibitem

\bibitem{CIFAR-10}
\begin{botherref}
\oauthor{\bsnm{Krizhevsky}, \binits{A.}},
\oauthor{\bsnm{Hinton}, \binits{G.}}, et al.:
Learning multiple layers of features from tiny images
(2009)
\end{botherref}
\endbibitem

\bibitem{RESNET20}
\begin{bchapter}
\bauthor{\bsnm{He}, \binits{K.}},
\bauthor{\bsnm{Zhang}, \binits{X.}},
\bauthor{\bsnm{Ren}, \binits{S.}},
\bauthor{\bsnm{Sun}, \binits{J.}}:
\bctitle{Deep residual learning for image recognition}.
In: \bbtitle{CVPR}
(\byear{2016})
\end{bchapter}
\endbibitem

\bibitem{TinyTaxiNet}
\begin{bchapter}
\bauthor{\bsnm{Julian}, \binits{K.D.}},
\bauthor{\bsnm{Lee}, \binits{R.}},
\bauthor{\bsnm{Kochenderfer}, \binits{M.J.}}:
\bctitle{Validation of image-based neural network controllers through adaptive
  stress testing}.
In: \bbtitle{ITSC}
(\byear{2020})
\end{bchapter}
\endbibitem

\bibitem{szegedy2013intriguing}
\begin{botherref}
\oauthor{\bsnm{Szegedy}, \binits{C.}},
\oauthor{\bsnm{Zaremba}, \binits{W.}},
\oauthor{\bsnm{Sutskever}, \binits{I.}},
\oauthor{\bsnm{Bruna}, \binits{J.}},
\oauthor{\bsnm{Erhan}, \binits{D.}},
\oauthor{\bsnm{Goodfellow}, \binits{I.}},
\oauthor{\bsnm{Fergus}, \binits{R.}}:
Intriguing properties of neural networks.
arXiv preprint arXiv:1312.6199
(2013)
\end{botherref}
\endbibitem

\bibitem{badnets}
\begin{barticle}
\bauthor{\bsnm{Gu}, \binits{T.}},
\bauthor{\bsnm{Liu}, \binits{K.}},
\bauthor{\bsnm{Dolan-Gavitt}, \binits{B.}},
\bauthor{\bsnm{Garg}, \binits{S.}}:
\batitle{Badnets: Evaluating backdooring attacks on deep neural networks}.
\bjtitle{IEEE Access}
\bvolume{7},
\bfpage{47230}--\blpage{47244}
(\byear{2019}).
\doiurl{10.1109/ACCESS.2019.2909068}
\end{barticle}
\endbibitem

\bibitem{DBLP:conf/kbse/GopinathCPT19}
\begin{bchapter}
\bauthor{\bsnm{Gopinath}, \binits{D.}},
\bauthor{\bsnm{Converse}, \binits{H.}},
\bauthor{\bsnm{Pasareanu}, \binits{C.S.}},
\bauthor{\bsnm{Taly}, \binits{A.}}:
\bctitle{Property inference for deep neural networks}.
In: \bbtitle{34th {IEEE/ACM} International Conference on Automated Software
  Engineering, {ASE} 2019, San Diego, CA, USA, November 11-15, 2019},
pp. \bfpage{797}--\blpage{809}.
\bpublisher{{IEEE}}, \blocation{???}
(\byear{2019}).
\doiurl{10.1109/ASE.2019.00079}.
\burl{https://doi.org/10.1109/ASE.2019.00079}
\end{bchapter}
\endbibitem

\end{thebibliography}

\end{document}